\documentclass[fleqn,usenatbib]{mnras}

\usepackage[T1]{fontenc}

\DeclareRobustCommand{\VAN}[3]{#2}
\let\VANthebibliography\thebibliography
\def\thebibliography{\DeclareRobustCommand{\VAN}[3]{##3}\VANthebibliography}

\usepackage{graphicx}	
\usepackage{amsmath}	
\usepackage{amssymb}	
\usepackage{newtxtext,newtxmath}
\usepackage{booktabs}
\usepackage{tabularx}
\usepackage{makecell}
\usepackage{multicol}
\usepackage{changepage}

\title[Benchmarking PhotoZ Methods]{Benchmarking and Scalability of Machine Learning Methods for Photometric Redshift Estimation}

\author[B. Henghes et al.]{
Ben Henghes,$^{1}$\thanks{E-mail: ben.henghes.13@ucl.ac.uk}
Connor Pettitt,$^{2}$
Jeyan Thiyagalingam,$^{2}$\thanks{E-mail: t.jeyan@stfc.ac.uk}
Tony Hey,$^{2}$
and Ofer Lahav $^{1}$
\\
$^{1}$University College London, Gower Street, London, WC1E 6BT, UK \\
$^{2}$Scientific Computing Department, Rutherford Appleton Laboratory, Science and Technology Facilities Council (STFC), Harwell Campus, Didcot, OX11 0QX, UK \\
}

\date{Accepted XXX. Received YYY; in original form ZZZ}

\pubyear{2021}

\begin{document}
\label{firstpage}
\pagerange{\pageref{firstpage}--\pageref{lastpage}}
\maketitle

\begin{abstract}
Obtaining accurate photometric redshift estimations is an important aspect of cosmology, remaining a prerequisite of many analyses. In creating novel methods to produce redshift estimations, there has been a shift towards using machine learning techniques. However, there has not been as much of a focus on how well different machine learning methods scale or perform with the ever-increasing amounts of data being produced. Here, we introduce a benchmark designed to analyse the performance and scalability of different supervised machine learning methods for photometric redshift estimation. Making use of the Sloan Digital Sky Survey (SDSS - DR12) dataset, we analysed a variety of the most used machine learning algorithms. By scaling the number of galaxies used to train and test the algorithms up to one million, we obtained several metrics demonstrating the algorithms’ performance and scalability for this task. Furthermore, by introducing a new optimisation method, time-considered optimisation, we were able to demonstrate how a small concession of error can allow for a great improvement in efficiency. From the algorithms tested we found that the Random Forest performed best in terms of error with a mean squared error, $MSE = 0.0042$; however, as other algorithms such as Boosted Decision Trees and k-Nearest Neighbours performed incredibly similarly, we used our benchmarks to demonstrate how different algorithms could be superior in different scenarios. We believe benchmarks such as this will become even more vital with upcoming surveys, such as LSST, which will capture billions of galaxies requiring photometric redshifts.
\end{abstract}




\section{Introduction}

Calculating distances to cosmological objects remains one of the most important steps required for probing cosmology. These distances are given by the distance-redshift relation, and hence one needs very accurate measures of redshift to be confident in the inferred distances. Ideally, high resolution spectra would be obtained for every object enabling for a precise measurement of the redshift. However, with current and future surveys such as the Dark Energy Survey (DES) \citep{dark2005dark, DES_more}, Euclid \citep{amendola2018cosmology}, and the Vera C. Rubin Observatory's Legacy Survey of Space and Time (LSST) \citep{tyson2003lsst, Ivezi__2019}, even with large spectroscopic surveys such as the Dark Energy Spectroscopic Instrument (DESI) \citep{flaugher2014dark, desi_2}, only tens of millions of the galaxies will have spectroscopy performed, despite hundreds of millions of galaxies being observed. 

In the absence of real spectroscopic measurements, estimating the photometric redshifts (Photo-Z) is the only viable route available for scientists. There are two major techniques used for photometric redshift estimation, template flitting (eg. \citet{benitez2000bayesian}) and machine learning (ML) (eg. \citet{collister2004annz}). Both methods rely on the photometric information produced by the survey, usually given as magnitudes in different colour bands. These magnitudes act as approximate measures of the underlying spectral energy distribution (SED) of the observed object, and by appropriately reconstructing the SED, a corresponding redshift can be inferred \citep{bolzonella2000photometric}.  

Template fitting methods use a small and fixed set of template spectra for the estimations, and inherently relies on the assumption that the best fitting SED template provides the true representation of the observed SED. There are benefits of template methods, such as, the ability to incorporate physical information, like dust extinction, into the model. However, embedding such physical constraints require very precise calibration and an accurate model~\citep{benitez2000bayesian}. 

Machine learning techniques, on the other hand, do not have any explicit model for capturing the physical information of the objects or of the estimation process. Instead, ML techniques rely on a training dataset with spectroscopic redshifts from observed or simulated (or a combination of both) data for inferring an estimation model. More specifically, supervised learning models rely on a guided principle, that with sufficient examples of input-output pairs an estimation model can be inferred by understanding the latent variables of the process. In other words, ML methods derive a suitable functional mapping between the photometric observations and the corresponding redshifts.

The learning process relies on a labelled dataset consisting of a set of magnitudes in each wavelength band (the inputs) and corresponding true values of the spectroscopic redshifts (the output labels or ground-truth). The learning model, such as a random forest or neural network, learns the mappings which can be non-linear. It has been shown that the functional mapping learned through the supervised learning outperforms the template-based methods~\citep{abdalla2011comparison}. 

Although the usage of ML in handling this problem has become very common ~\citep{pasquet2019photometric, d2018photometric, hoyle2016measuring}, there is still no comprehensive study outlining the overall understanding on different ML methods in handling the Photo-Z problem. In fact, this is a common problem across all domains of sciences, and as such, the notion of AI benchmarking is an upcoming challenge for the AI and scientific community. This is particularly true in light of recent developments in the ML and AI domains, such as the deep learning revolution~\citep{sejnowski2018deep}, technological development on surveys~\citep{dewdney2009square}, the ability to generate or simulate synthetic data~\citep{springel2005cosmological}, and finally the progress in computer architecture space, such as the emergence of GPUs~\citep{kirk2007nvidia}. 

The notion of benchmarking~\citep{dongarra2003linpack} has conventionally been about how a given architecture (or an aspect of a given architecture) performs for a given problem, such as the LINPACK challenge~\citep{dongarra1979linpack}. However, in our case, the focus is broader than just performance. Our motivation here is many-fold, including understanding how different ML models compare when estimating the redshifts, how these techniques perform  when the available training data is scaled, and finally how these techniques scale for inference. Furthermore, one of the key challenges here is the identification of appropriate metrics or figures of merit for comparing these models across different cases. 

We intend to answer some of these questions in this paper by introducing this as a representative AI  benchmarking problem from the astronomical community. The benchmarks will include several baseline reference implementations covering different ML models and address the challenges outlined above. The rest of this paper is organised as follows: In Section~\ref{sec:data}, we describe the the dataset used and include discussions on the features selected. In Section~\ref{sec:method}, we briefly describe the machine learning models that were evaluated in the study, followed by the descriptions of the optimisation and benchmarking processes and the different metrics that are part of our analyses. The results are then presented in Section~\ref{sec:results} along with our observations, and we conclude the paper in Section~\ref{sec:conclusion} with directions for further work.

\section{Data}
\label{sec:data} 

The data used in our analysis comes entirely from the Sloan Digital Sky Survey (SDSS) \citep{york2000sloan}. Using its dedicated 2.5 meter telescope at Apache Point Observatory \citep{gunn20062}, SDSS is one of the largest public surveys with over 200 million photometric galaxies and 3 million useful galaxy spectra as of data release 12 (DR12) \citep{alam2015eleventh}. 

In this work we downloaded 1,639,348 of these galaxies with spectroscopic data available to be used by the machine learning algorithms. The spectroscopic redshift was required as it was taken to be the ground truth for the redshift that the algorithms were trying to predict using the magnitudes of each galaxy. SDSS took images using five different optical filters (u, g, r, i, z), and as a result of these different wavelength bands, there were five magnitudes for each observed galaxy \citep{eisenstein2011sdss}. 

The 1.6 million galaxies used in this investigation were from a cleaned dataset where it was a requirement for all five magnitudes to have been measured. In many cases for observations of galaxies there could be a missing value in one of the filters which would negatively impact it's redshift estimation. By only using galaxies with complete photometry we ensured that our comparison of methods wasn't also being affected by the kind of galaxies within the different sized datasets.

Furthermore, the redshift range of the galaxies used was constrained to only have galaxies with a redshift, $z < 1$. As there are far fewer galaxies with measured spectroscopic redshifts greater than 1, we kept within this range to ensure that the training set would be representative and allow for reliable estimates to be generated. This meant that the benchmarking performed could be carried out without also having to take into account the effects that an unclean dataset might have had on the different machine learning algorithms. 

The main features of the data used by machine learning algorithms were the five magnitudes which could also be combined to give the four colours that are simply the difference in magnitudes between neighbouring wavelength bands (u-g, g-r, r-i, and i-z). There were additional feature columns contained in the SDSS data which could have been added such as the subclass of galaxy or the Petrosian radius \citep{petrosian1976surface, Soo}. However; adding these additional features wouldn't have had a large impact on the results and could have added more issues due to incompleteness if the feature wasn't recorded for every galaxy. Instead it was decided to use only the information from the five magnitudes as features which we knew to be complete. 

Finally, we also scaled the features by subtracting its mean and diving by its standard deviation to give unit variance. This ensured that the machine learning algorithms used weren't being influenced by the absolute size of the values, where a difference in a feature's variance could result in it being seen as more important than other features. And by randomly splitting the full dataset to form the training and testing sets, the subsets created kept the same distribution of redshift and were representative of the overall dataframe.

\section{Methodology} \label{sec:method}

With the data prepared, the first step of the machine learning process was to split the entire dataset to create a training set, testing set, and validation set, whereby the test and validation sets were kept unseen by the machine learning algorithms until after it had been trained using the training data. As part of the benchmarking process the machine learning algorithms (described in Sec~\ref{ml_descriptions}) were trained and tested on many different sizes of datasets, and to do this the data was split randomly for each size of training and testing set required. 

During training, the algorithms were also optimised by changing the hyperparameters. These are the parameters of the models that control how the machine learning algorithms create their mappings from the features to the redshift. The most complete way of optimising would be to perform brute force optimisation where every combination of a defined grid of hyperparameters would be tested. However, this is far more computationally intensive than random optimisation which instead tests a random subset of the hyperparameter grid and provides a good estimate of the best hyperparameters. The grids of hyperparameters tested for each algorithm is given in Table~\ref{tab:hyperparameters} along with the selected parameters.

\begin{table*}
\centering
\caption{Grids of hyperparameters that were searched to test and compare each machine learning algorithm, along with the hyperparameters that were selected by the random optimisation. The arrays of hyperparameters were chosen to give a good overview of different possible configurations of the algorithms, and by changing the parameters which had the greatest impact on the algorithms, we ensured finding a good representation of the `best' performing algorithms.}
	\label{tab:hyperparameters}
	\begin{tabular}{cccc}
		\hline
		 Classifier & Hyperparameter & Array of Values Searched & Selected Value \\
 		\hline
    	   LR & ``fit intercept" & [True, False] & \textbf{True}\\
              & ``normalize"& [True, False] & \textbf{True} \\
		\hline
		kNN & ``no. neighbors" & [1, 200] & \textbf{21}\\
             & ``weights" & [``uniform", ``distance"] & \textbf{``distance"}\\
             & ``leaf size" & [10, 100] & \textbf{27}\\
             & ``p" & [1, 4] & \textbf{2} \\
		\hline
		DT & `max. features" & [1, 5, ``auto"] & \textbf{``auto"}\\
		   & ``min. samples split" & [2, 100] & \textbf{38}\\
		   & ``min. samples leaf" & [1, 100] & \textbf{64}\\  
		   & ``min. weight fraction leaf" & [0, 0.4] & \textbf{0}\\
           & ``criterion" & [``mse", ``mae"] & \textbf{mse}\\
        \hline
		BDT & ``no. estimators" & [1, 200] & \textbf{88}\\
		    & ``loss" & [``ls", ``lad", ``huber", ``quantile"] & \textbf{``lad"}\\
		    & ``max. features" & [1, 5] & \textbf{4} \\
            & ``max. depth" & [1, 20] & \textbf{17} \\
		    & ``min. samples split" & [2, 100] & \textbf{46} \\
		    & ``min weight fraction leaf" & [0, 0.4] & \textbf{0} \\
        \hline
		RF & ``no. estimators" & [1, 200] & \textbf{94}\\
			& ``max. features" & [1, 5] & \textbf{4} \\
            & ``min. samples leaf" & [1, 100] & \textbf{8}\\
            & ``min. samples split" & [2, 100] & \textbf{13} \\
		    & ``min weight fraction leaf" & [0, 0.4] & \textbf{0} \\
		    & ``criterion" & [``mse", ``mae"] & \textbf{mae}\\
        \hline
		ERT & ``no. estimators" & [1, 200] & \textbf{147}\\
			& ``max. features" & [1, 5] & \textbf{4} \\
            & ``min. samples leaf" & [1, 100] & \textbf{3}\\
            & ``min. samples split" & [2, 100] & \textbf{87} \\
		    & ``min weight fraction leaf" & [0, 0.4] & \textbf{0} \\
		    & ``criterion" & [``mse", ``mae"] & \textbf{mse}\\
        \hline
		MLP & ``hidden layer sizes" & [(100, 100, 100), (100, 100), 100] & \textbf{(100, 100, 100)}\\
		    & ``activation" & [``tanh", ``relu"] & \textbf{``tanh"}\\
		    & ``solver" & [``sgd", ``adam"] & \textbf{``adam"}\\
		    & ``alpha" & [0.00001, 0.0001, 0.001, 0.01] & \textbf{0.01}\\
            & ``tol" & [0.00001, 0.0001, 0.001, 0.01] & \textbf{0.00001}\\
            & ``learning rate" & [``constant",``adaptive"] & \textbf{``constant"}\\
        \hline
	\end{tabular}
\end{table*}

To be able to optimise the algorithms the decision first had to be made of which metric would be optimised for. There are 3 main metrics used for regression problems such as this: mean squared error (MSE), mean absolute error (MAE), and R squared score ($R^2$). The formulae for calculating each of these metrics is given below where for the $i$-th sample within a total of $n$ samples, $\hat{z}_i$ is the predicted value, and $z_i$ is the true value.

\begin{equation} 
\text{MSE}(z, \hat{z}) = \frac{1}{n_{\text{samples}}} \sum_{i=0}^{n_{\text{samples}}-1} (z_i - \hat{z}_i)^2
\end{equation}

\begin{equation} 
\text{MAE}(z, \hat{z}) = \frac{1}{n_{\text{samples}}} \sum_{i=0}^{n_{\text{samples}}-1} \left| z_i - \hat{z}_i \right|
\end{equation}

\begin{equation} 
R^2(z, \hat{z}) = 1 - \frac{\sum_{i=1}^{n} (z_i - \hat{z}_i)^2}{\sum_{i=1}^{n} (z_i - \bar{z})^2}
\end{equation}

There are three additional metrics defined below that are commonly used to determine the performance of photometric redshift estimations: bias (the average separation between prediction and true value), precision (also $1.48 \times$ median absolute deviation (MAD) which gives the expected scatter), and outlier fraction (the fraction of predictions where the error is greater than a set threshold, here chosen to be $>0.10$). Each of these metrics were also calculated and the results are given in Table~\ref{tab:mag_results}.

\begin{equation} 
\text{Bias} = <z_{\text{pred}} - z_{\text{spec}}>
\end{equation}

\begin{equation} 
\text{Precision} = 1.48 \times \text{median}(\frac{|z_{\text{pred}} - z_{\text{spec}}|}{1 + z_{\text{spec}}}) 
\end{equation}

\begin{equation} 
\text{Outlier Fraction} = \frac{N(\Delta z) > 0.10}{N_{\text{total}}}
\end{equation}

As well as deciding which metric to optimise for, we introduced an extra stage included in the optimisation which allowed for a time-considered optimisation (see section \ref{time-considered optimisation}). We optimised the machine learning algorithms for MSE (aiming to minimise the MSE) and used a random optimisation with 1000 iterations to ensure a good estimate of the best hyperparameters for each algorithm. Furthermore, we used a 3-fold cross validation \citep{breiman1992submodel} to ensure that the algorithms weren't overfitting (which could mean that the algorithms were able to perform well for the training data used but then fail to generalise), and that the results would be valid for any given dataset. Once optimised each algorithm was then retrained and tested to give the final results given in Sec \ref{sec:results}, along with the benchmarking results, where the benchmarking process used is described in \ref{benchmarking}.

\subsection{Descriptions of Machine Learning Algorithms Tested}
\label{ml_descriptions}

The following algorithms were selected for testing as they are some of the most widely used machine learning algorithms and all available though the python package Scikit-Learn \citep{pedregosa2011scikit}. While a simple neural network (Multi-layer Perceptron) was included, we didn't include any other examples of deep learning. This decision was made as deep learning algorithms perform best with many features (often thousands) and there's only so much information that the photometry could provide with the five magnitude features. Furthermore, it's been shown by \cite{hoyle2016measuring} that ``traditional" algorithms can perform equally well as deep learning methods, and that it might only be beneficial to use more computationally expensive deep learning models when directly using images as the training data \citep{pasquet2019photometric}. 

\subsubsection{Linear Regression}

Linear Regression, or Ordinary Least Squares Regression, fits a linear model to the data with coefficients that act to minimize the sum of the squared residuals between the observations and the predictions from the linear approximation. The linear model requires independent features, as features that are correlated will give estimates that are very sensitive to random errors in the observations, resulting in a large variance \citep{hastie2009linear}.

\subsubsection{K-Nearest Neighbours}

K-Nearest Neighbours uses a predefined number of data points in the training sample, k, which are closest in Euclidean distance to the new point whose value is then predicted based off those. This is an example of an instance based algorithm, where there is no general model used to make predictions but which stores the training data. Although one of the simplest methods, being non-parametric can make it very successful especially in cases with an irregular decision boundary. Increasing the value of k acts to reduce the effects of noise, however, it also makes the decision boundary less distinct and could result in overfitting \citep{knn}.

\subsubsection{Decision Trees}

Decision Trees \citep{Breiman1983ClassificationAR} are non-parametric algorithms whereby the data features are used to learn simple decision rules. The decision rules are basic if-then-else statements and are used to split the data into branches. The tree is then trained by recursively selecting the best feature split which is taken to be the split which gives the highest information gain, or greatest discrepancy between the two classes. Typically Decision trees can produce results with a high accuracy, however, they are bad at generalising the data as they are often complex and overfitted. Instead they can be combined in an ensemble such as Boosted Decision Trees, Random Forests, or Extremely Randomised Trees.

\subsubsection{Boosted Decision Trees}

Boosted Decision Trees were the first ensemble method we considered. The process of boosting can be described whereby the machine learning algorithm is repeatedly fitted to the same dataset, but each time the weights of objects with higher errors are increased. This aims to result in an algorithm that can better handle the less common cases than a standard decision tree. The boosting can be generalised by using an arbitrary differentiable loss function which is then optimised \citep{BDT1}, and we found that for this problem using the least absolute deviation loss function produced the best results.

\subsubsection{Random Forests}

Random Forests also take many decision trees to build an ensemble method, averaging the predictions of the individual trees to result in a model with lower variance than in the case of a single decision tree. This is done by adding two elements of randomness, the first of which is using a random subset of the training data which is sampled with replacement \citep{Bagging}. Second, the feature splits are found from a random subset of the features rather than using the split which results in the greatest information gain. This randomness can yield decision trees with higher errors, however, by averaging the predictions, the errors cancel out and the variance reduction yields a greatly improved model as well as removing the typical overfitting that occurs with single decision trees \citep{RF1}.

\subsubsection{Extremely Randomised Trees}

Extremely Randomised Trees \citep{ERT} is an algorithm very similar to Random Forests but with an additional step to increase randomness. The feature splits are not only found from a random subset of the features. It also uses thresholds that are picked at random for each feature before the best of these random thresholds are then used for the decision rules, instead of simply using the thresholds which results in the greatest information gain. This acts to further reduce the variance compared to a Random Forest, however, it also results in a slightly greater bias.

\subsubsection{Multi-layer Perceptron}

The Multi-layer Perceptron is the most simple example of a fully connected, deep neural network with at least three layers of nodes. It consists of the input node, output node, and a minimum of one hidden layer, although more can be added. In the way that the perceptron learns how to map the input node to the target vector, it's similar to logistic regression, however, it differs with the addition of one or more non-linear hidden layers which allow it to approximate any continuous function \citep{WERBOS1988339, MLP}.

\begin{table*}
\centering
	\caption{Results of testing the seven machine learning algorithms described in Sec~\ref{ml_descriptions}. Each algorithm was trained using 10000 galaxies and tested using 5-fold cross validation to obtain the quoted standard deviation.}
	\label{tab:mag_results}
	\begin{tabularx}{\textwidth}{XXXXXXXX}
	\toprule
	\thead{} & \thead{Linear \\ Regression \\ (LR)} & \thead{k-Nearest \\Neighbours \\ (kNN)} &  \thead{Decision \\ Tree \\ (DT)} & \thead{Boosted \\ Decision Tree \\ (BDT)} &  \thead{Random \\ Forest \\ (RF)} & \thead{Extremely\\ Randomised \\ Trees (ERT)} & \thead{Multi-layer \\ Perceptron \\ (MLP)} \\
	\midrule
	\addlinespace[0.2cm]
	MSE & 0.005714 $\pm$0.000577 & 0.004438  $\pm$0.000417 & 0.004631  $\pm$0.000407 & 0.004277 $\pm$0.000394 & 0.004221 $\pm$0.000423 & 0.004327 $\pm$0.000419 & 0.004701 $\pm$0.000499   \\
	\addlinespace[0.2cm]
	\midrule
	\addlinespace[0.2cm]
	MAE & 0.050931 $\pm$0.001679 & 0.040881 $\pm$0.001626 & 0.041827 $\pm$0.001452 & 0.038757 $\pm$0.001514 & 0.038504 $\pm$0.001484 & 0.040459 $\pm$0.001537 & 0.051260 $\pm$0.008874  \\
	\addlinespace[0.2cm]
	\midrule
	\addlinespace[0.2cm]
	$R^2$ & 0.865198 $\pm$0.009009 & 0.895208 $\pm$0.007215 & 0.890677 $\pm$0.006415 & 0.899017 $\pm$0.006822 & 0.900366 $\pm$0.007373 & 0.897861 $\pm$0.007198 & 0.871507 $\pm$0.014329 \\
	\addlinespace[0.2cm]
	\midrule
	\addlinespace[0.2cm]
 	Bias & 0.039742 $\pm$0.000920 & 0.031109 $\pm$0.000977 & 0.032030 $\pm$0.000895 & 0.029428 $\pm$0.000893 & 0.029334 $\pm$0.000854 & 0.030927 $\pm$0.000947 & 0.034577 $\pm$0.002209\\
 	\addlinespace[0.2cm]
 	\midrule
 	\addlinespace[0.2cm]
 	Precision & 0.043421 $\pm$0.000578 & 0.031836 $\pm$0.000845 & 0.032895 $\pm$0.001137 & 0.028986 $\pm$0.000264 & 0.029279 $\pm$0.000766 & 0.031945 $\pm$0.000609 & 0.040837 $\pm$0.003310 \\
 	\addlinespace[0.2cm]
 	\midrule
 	\addlinespace[0.2cm]
 	Outlier Fraction & 0.060500 $\pm$0.005187 & 0.034800 $\pm$0.007033 & 0.033400 $\pm$0.007439 & 0.029300 $\pm$0.005183 & 0.029700 $\pm$0.004389 & 0.033400 $\pm$0.006304 & 0.037600 $\pm$0.002709 \\
 	\addlinespace[0.2cm]
 	\bottomrule
	\end{tabularx}
\end{table*}

\subsection{Time-Considered Optimisation} \label{time-considered optimisation}

In the normal process of optimising machine learning algorithms, a single metric is chosen to minimise. If brute force optimisation is used, this produces an algorithm configured with the hyperparameters from the defined grid which gives the best result for the metric (e.g. the lowest MSE). Although this algorithm by definition would have the best result, it doesn't necessarily result in the most useful or suitable algorithm. The hyperparameters selected to minimise the error likely also act to increase the computational time required both in training and inference, resulting in a much slower model. 

Rather than minimising a single metric, in time-considered optimisation we also consider the time taken by the models both in training and inference. By setting an error tolerance we allow for the model selection to suggest an alternative to the `best' model (the model which minimises the error metric), instead providing a model which will have a higher error, while kept below the tolerance, but in return will also have faster training and inference times. In certain cases, such as training the Decision Trees, it was possible to achieve a two magnitude increase in efficiency while increasing the error by $< 10\%$. 

For the purpose of benchmarking the machine learning algorithms in this paper, we set the error tolerance to machine precision (which is usually  $10^{-16}$, close to zero) resulting in the `best' model in terms of error. This decision was made as these optimised algorithms would result in the algorithms most commonly used in other machine learning studies where time-considered optimisation hasn't be implemented.

\subsection{Benchmarking} \label{benchmarking}

The benchmarking performed was achieved by recording the system state (described by the time, CPU usage, memory usage, and disk I/O) throughout the process of running the machine learning algorithms. This allowed us to compare the efficiency of both training and inferencing performance of the machine learning models, and when combined with the regression errors obtained, allowed for a complete description of the performance of the different methods. 

\begin{figure*}
\begin{multicols}{2}
    \begin{adjustwidth}{-0.7cm}{}
    \includegraphics[scale=0.34]{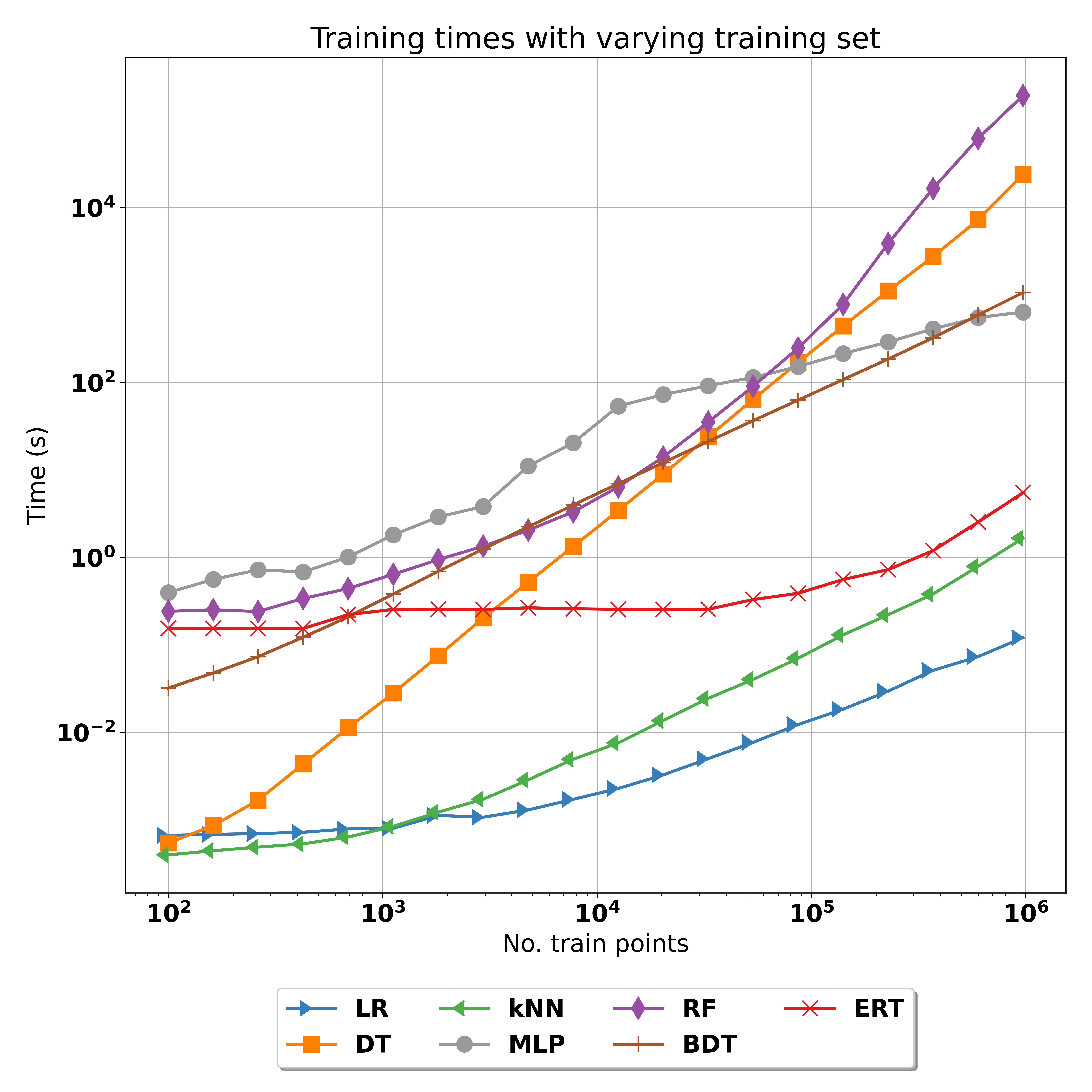}\par 
    \end{adjustwidth}
        \caption{Graph of the training time plotted against the number of galaxies used in the training set to show how each algorithm scales with different sizes of training datasets. 
        We saw that the simpler LR, kNN, and DT algorithms all begin as the fastest to train, however, the DT had terrible scaling and for large training sets became one of the slowest algorithms. Conversely the ERT and MLP algorithms began as two of the slowest algorithm to train, but scaled much better than the rest and could be more useful for massive training datasets.}
        \label{fig:benchmarks_training_times}
    \includegraphics[scale=0.34]{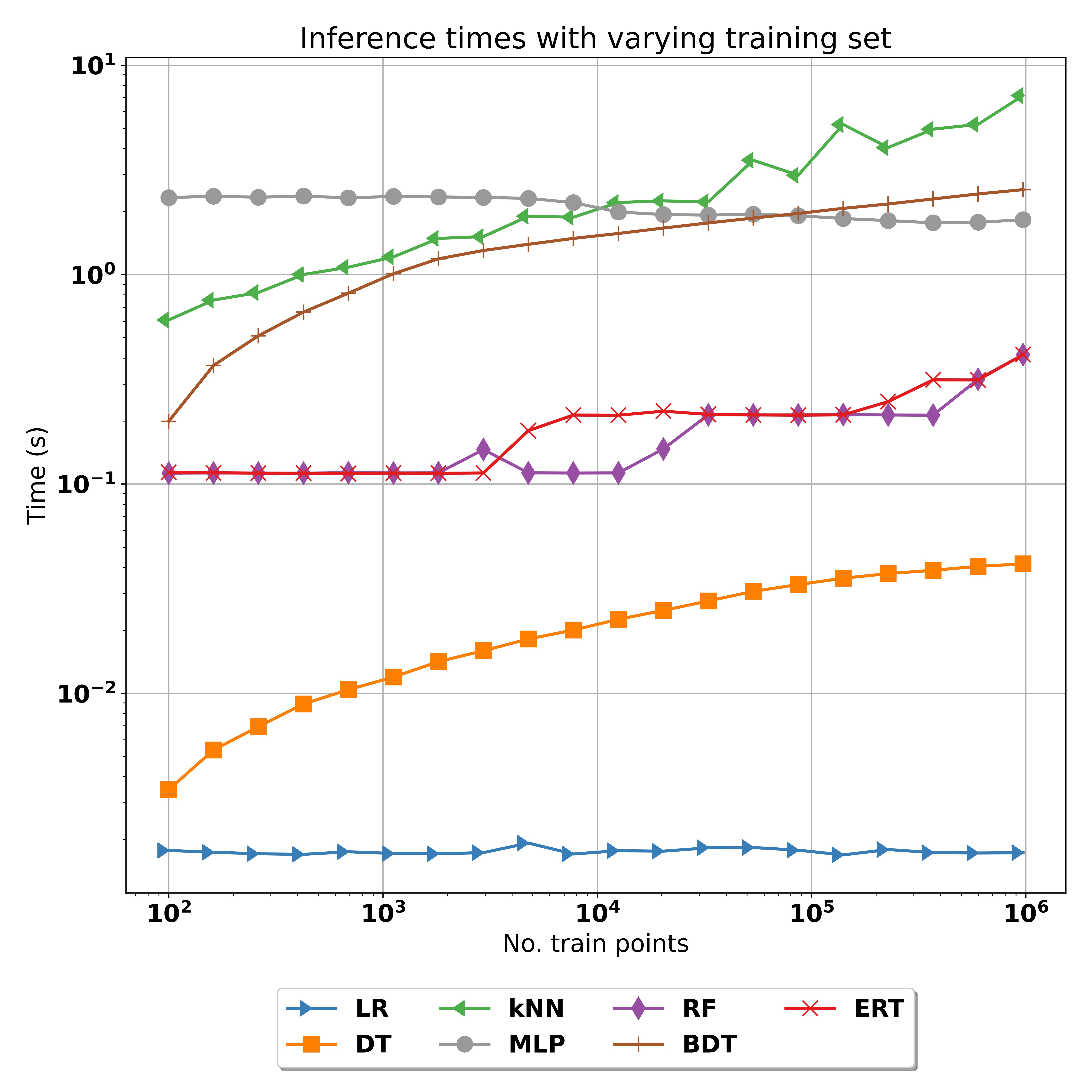}\par 
        \caption{Graph of the inference time plotted against the number of galaxies used in the training set to show how each algorithm scaled with different sizes of training datasets (and a constant test set of 327870 galaxies).
        We saw all algorithms other than LR and MLP exhibit a training bloat, whereby the inference time increased with the number of galaxies included in the training set; however, the algorithms inference times generally increased by only a factor of $10$ despite the training dataset increasing by a factor of $10^4$.}
        \label{fig:benchmarks_inference_times_training}
    \end{multicols}
\begin{multicols}{2}
    \begin{adjustwidth}{-0.7cm}{}
    \includegraphics[scale=0.34]{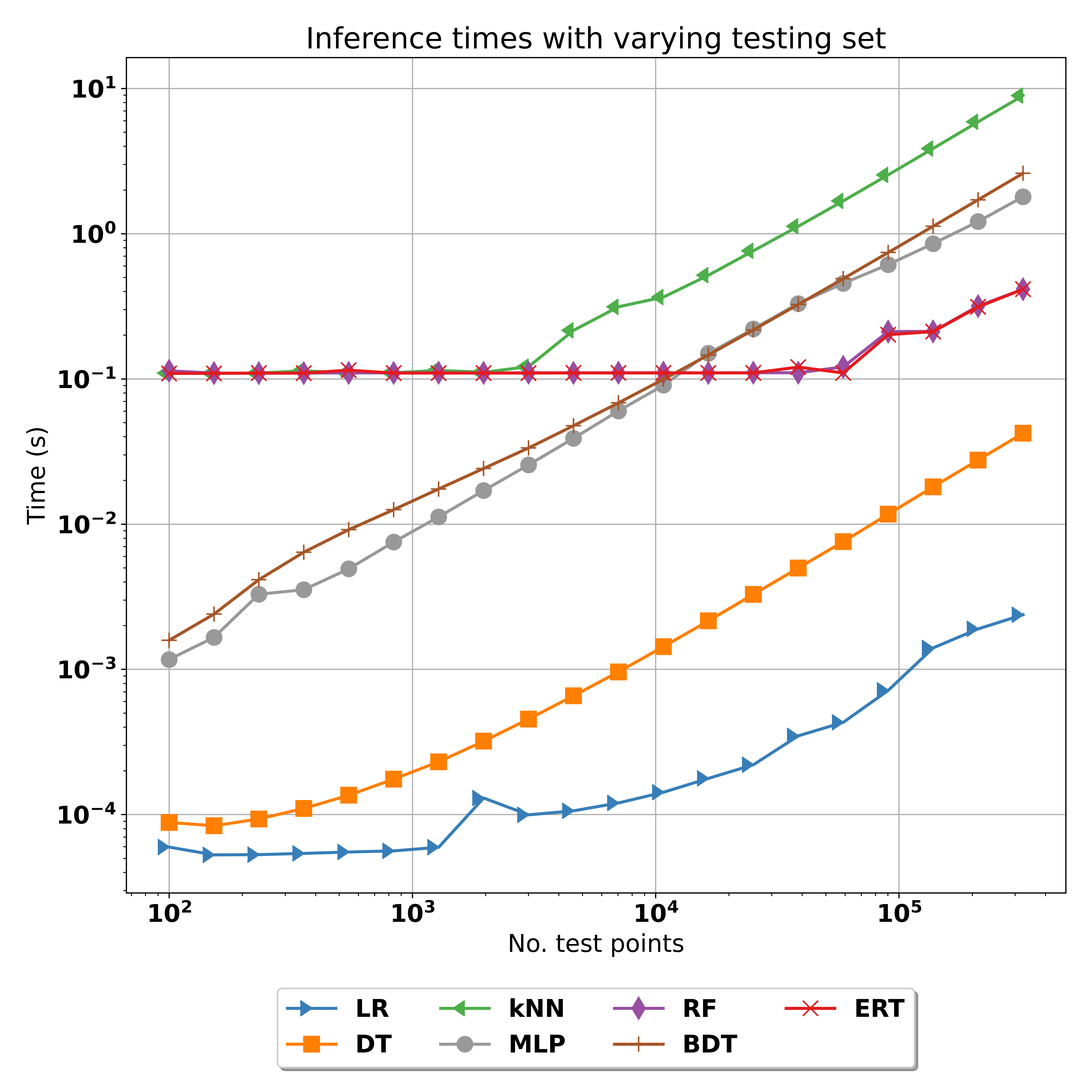}\par
    \end{adjustwidth}
        \caption{Graphs of the inference time plotted against the number of galaxies used in the testing set to show how each algorithm will scale with different sizes of testing datasets (and a constant training set of 983608 galaxies).
        In inference we saw all algorithms scaling very similarly with the main difference being the RF and ERT where, during the period between $10^2$ to $10^5$ galaxies used in the test set, the inference time didn't increase despite the number of galaxies to provide an estimate for increasing by a factor of $10^3$. This meant that both algorithms ended up being faster to provide redshift estimations for larger test sets.}
        \label{fig:benchmarks_inference_times_testing}
    \includegraphics[scale=0.342]{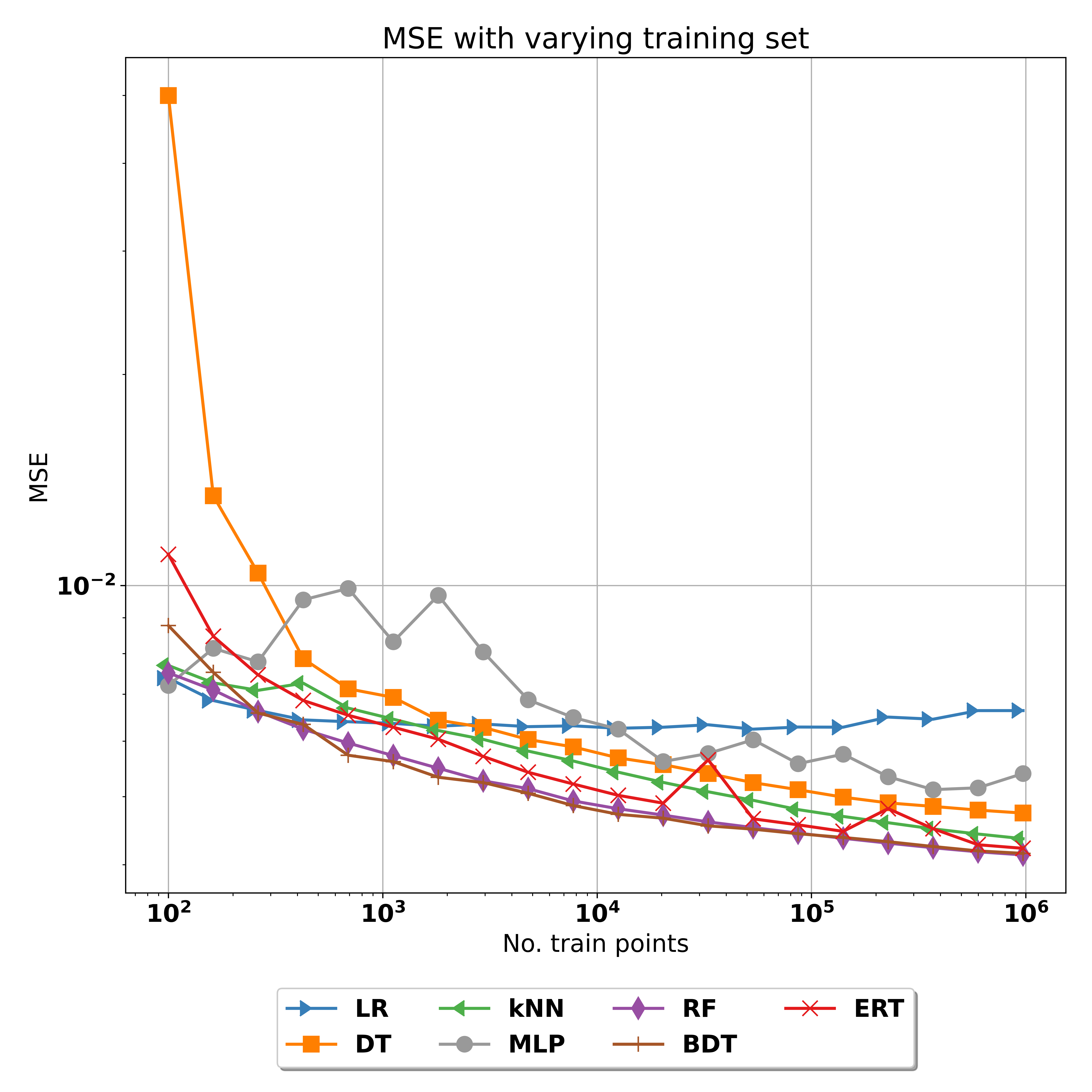}\par
        \caption{Graphs of the Mean Squared Error (MSE) plotted against the number of galaxies used in the training set to show how each algorithm's performance will scale with different sizes of training datasets (and a constant test set of 327870 galaxies). 
        As expected, in general we saw all algorithms (other than LR) achieving lower MSE as the number of galaxies included in the training set was increased. However, we saw this increased error performance quickly plateau, and past $10^4$ galaxies in the training set there was very little reduction in error. }
        \label{fig:benchmarks_mse}
\end{multicols}
\end{figure*}

Our main focus of the benchmark was to investigate how training and testing times varied with different sizes of dataframes, and how the final redshift estimations would be affected. As such we incrementally changed both the training and testing datasets and recorded the times taken which allowed us to produce the plots shown in Figures \ref{fig:benchmarks_training_times} - \ref{fig:benchmarks_mse}.

\section{Results} \label{sec:results}

\begin{figure*}
    \centering
    \includegraphics[scale = 0.6]{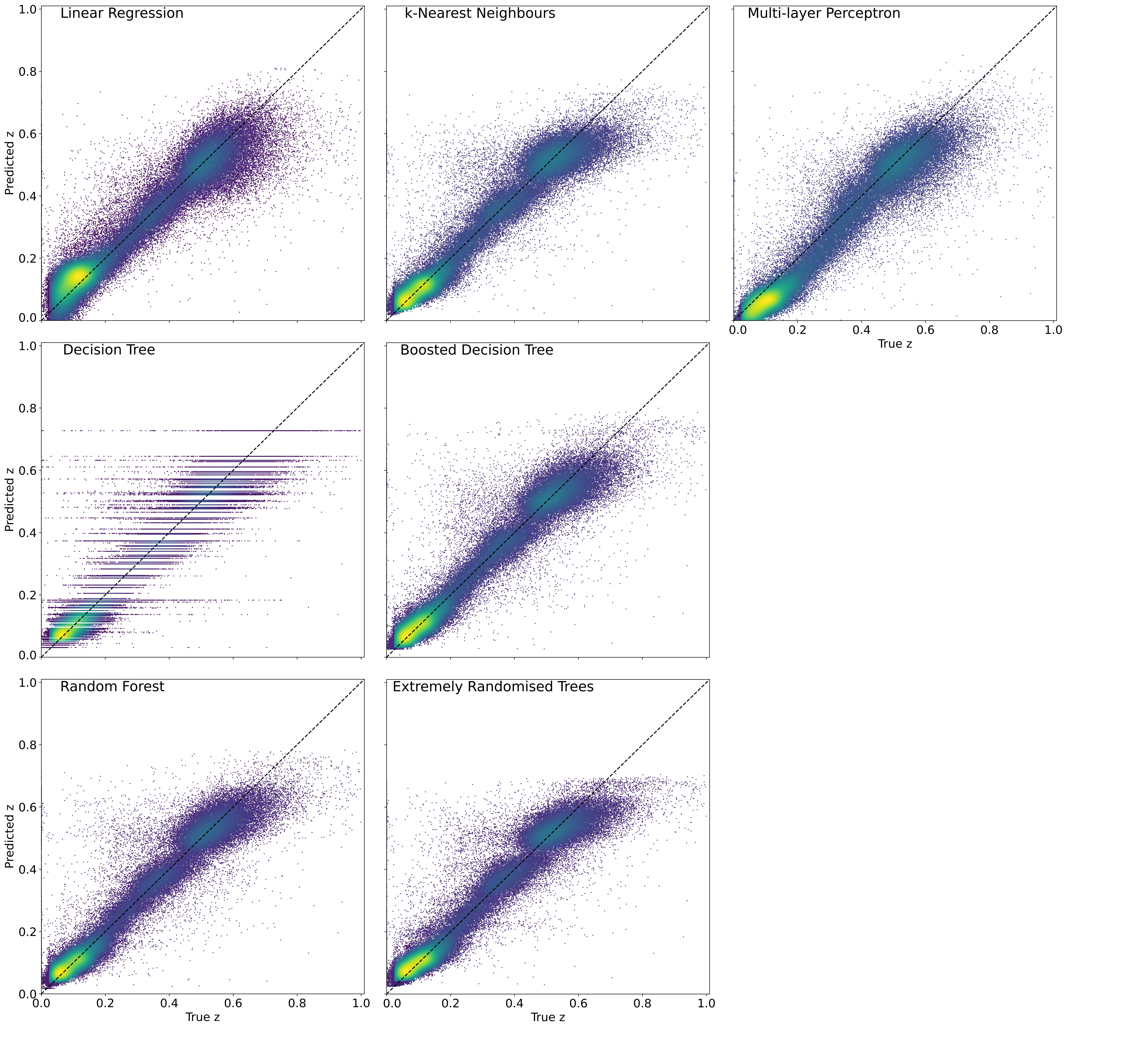}
    \caption{Graphs of photometric redshift estimates against the true spectroscopic redshift where the lighter shaded contours display the more densely populated regions. From top left to bottom  right - Linear Regression (LR), k-Nearest Neighbours (kNN), Multi-layer Perceptron (MLP), Decision Tree (DT), Boosted Decision Tree (BDT), Random Forest (RF), and Extremely Randomised Trees (ERT).}
    \label{fig:redshifts}
\end{figure*}

\begin{figure*}
    \centering
    \includegraphics[scale = 0.45]{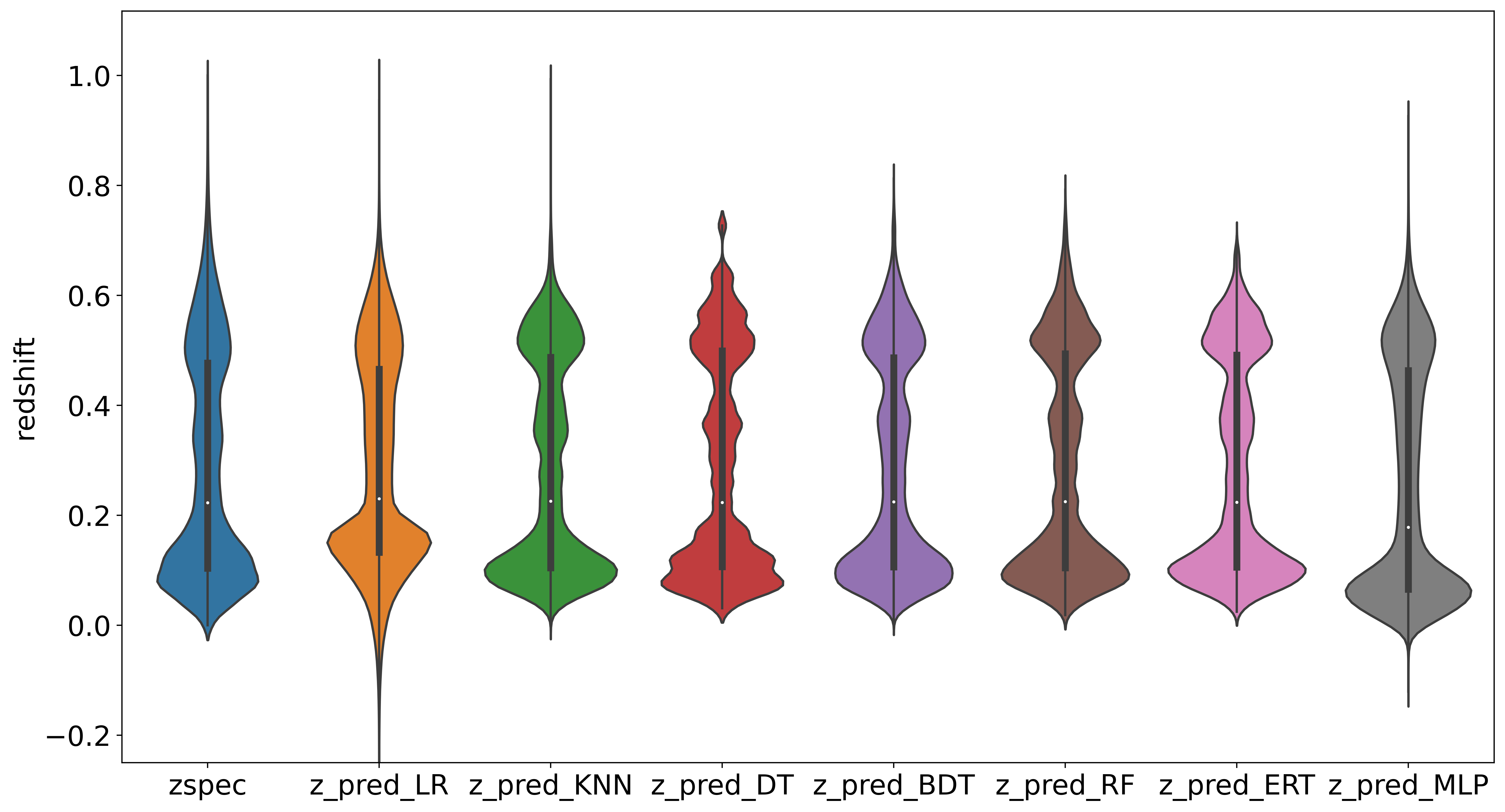}
    \caption{Violin plots showing the kernel density estimation of the underlying distributions of photometric redshift estimates of each algorithm along with the true spectroscopic redshift. From left to right - True spectroscopic redshift (zspec), Linear Regression (LR), k-Nearest Neighbours (kNN), Decision Tree (DT), Boosted Decision Tree (BDT), Random Forest (RF), Extremely Randomised Trees (ERT), and Multi-layer Perceptron (MLP).}
    \label{fig:violin}
\end{figure*}

The results given in Table~\ref{tab:mag_results} show how the seven machine learning algorithms performed at producing photometric redshift estimations. Furthermore Figure \ref{fig:redshifts} displays the true spectroscopic redshifts plotted against the photometric redshift estimates for each machine learning algorithm. We also plotted the distributions of the redshift estimations for each of the algorithms as well as the true spectroscopic redshift in a violin plot in Figure~\ref{fig:violin} to quickly see which algorithms were able to capture the correct distribution.

From these results we saw that all algorithms were able to successfully provide photometric redshift estimations. Using the violin plots from Figure~\ref{fig:violin} we could see that the rough distribution was recovered by each algorithm, with the Multi-layer Perceptron (MLP) producing a slightly more similar shape to the true redshifts. However, from simply looking at the outputs shown in Figures~\ref{fig:redshifts} \& \ref{fig:violin} it would be very difficult to determine which algorithm would be best to use. While the Decision Tree (DT) might be excluded due to the estimates being put into bands at set redshifts, its errors were still found to be quite low and it outperformed both the Linear Regression (LR) and MLP algorithms. 

Looking at the metrics in Table~\ref{tab:mag_results} alone, the Random Forest (RF) performed best having the lowest errors with a mean absolute error $(MAE) = 0.0385$ and mean squared error $(MSE) = 0.0042$, however, the other algorithms k-Nearest Neighbours (kNN), Boosted Decision Tree (BDT), and Extremely Randomised Trees (ERT) all performed incredibly similarly with $MAE < 0.042$ and $MSE < 0.0046$. Indeed, the BDT was almost identically performing to the RF with a slightly improved precision and outlier fraction, and with such close performances of all the other algorithms, it was impossible to sufficiently determine which algorithm would be the most useful. To be able to further differentiate between them and determine which would be the best algorithm to use, it was therefore necessary to use the benchmarking results. 

The results of the benchmarking performed for each algorithm are plotted in Figure~\ref{fig:benchmarks_training_times} (that shows the speed of training with varying sizes of training datasets), Figures~\ref{fig:benchmarks_inference_times_training} \& \ref{fig:benchmarks_inference_times_testing} (that show the inference speeds with varying sizes of either training or testing datasets), and Figure~\ref{fig:benchmarks_mse} (that shows how the MSE varies as the number of galaxies in the training set increases). As shown by these figures, the fastest algorithm overall was LR which remained the fastest both in training and inference with increasing sizes of training and testing datasets. This was perhaps not surprising as out of the algorithms tested it was the most simple model and as such required less computational resources both to train the model and to make its predictions. However, as LR also had by far the worst errors out of the algorithms tested (with errors around $30\%$ higher than those of the better performing algorithms), it seemed unlikely that it would ever be implemented for the problem of photometric redshift estimation.

Out of the other algorithms the DT and MLP were the poorer performing in terms of error. The DT was the second fastest behind LR in terms of inference, using its simple decision rules to quickly obtain the redshift estimations; however, as it also resulted in only estimating certain redshift bands, the final estimates weren't as useful as other algorithms. Furthermore, the DT was the worst scaling algorithm for training and became the second slowest algorithm to train on a million galaxies. The MLP was also one of the slowest algorithms tested, starting as the slowest to train with small training sets and also being one of the slowest in inference. Although it's the simplest example of deep learning, it suffered from being one of the more complex algorithms tested, and would perform better on even larger datasets with far more features, where it would have more chance to catch up to the other algorithms in both speed and error performance.

The remaining kNN, BDT, RF, and ERT algorithms all performed well in terms of error and were the hardest to differentiate between, however, using the benchmarking results it was possible to see how differently they scaled. kNN was the simplest of the four better performing algorithms, and using the nearest neighbours to produce it's estimates resulted in the second fastest training times, only being beaten by LR. Although kNN was very fast to train, it was the slowest in inference and exhibited a bad `training bloat' whereby the inference time increased as the number of galaxies in the training set was increased. While most other algorithms also displayed some level of this training bloat, it was worst for kNN due to the nature of it's nearest neighbour search which became more and more computationally expensive as more training points were added, and as such it wouldn't be as useful an algorithm for giving estimates for large datasets. 

Out of the three ensemble tree-based methods, the RF scaled the worst in terms of training, becoming the slowest algorithm to train on the 1 million galaxies. Whereas, the ERT scaled surprisingly well and became the third fastest algorithm in training and similar to kNN. In training the BDT was quite fast, scaling much better than the RF but worse than the ERT; however, when it came to inference the BDT scaled worse than both the ERT and RF and was the second slowest algorithm for large datasets. The RF and ERT scaled almost identically in inference, which made sense being such similar algorithms, both only being beaten by the much simpler LR and DT. 

As a result it seemed like there was no clear best performing algorithm, but rather each algorithm could be useful in different situations. While the RF had the best error metrics, its terrible scaling with increasing training data meant that it would only be the best algorithm for problems where it could be trained once and it would be inefficient to use for problems which required the algorithm to be regularly retrained on large amounts of data. In that case the BDT which had similar errors but was faster to train could be a more useful alternative, and similarly if the inference times were required to be lower the ERT would be a good compromise.

\section{Conclusions} \label{sec:conclusion}

Producing reliable photometric redshift estimations will continue to be an incredibly important area of cosmology, and with future surveys producing more data than ever before it will be vital to ensure that the methods chosen to produce the redshifts can be run efficiently. 

Here we showed how benchmarking can be used to provide a more complete view of how various machine learning algorithms' performances scale with differing sizes of training and testing datasets. By combining the benchmarking results with the regression metrics we were able to demonstrate how it's possible to distinguish between algorithms which appear to perform almost identically and suggest which could be better to implement in different scenarios. Furthermore, by suggesting a novel time-considered optimisation process which takes into account the benchmarking results during model selection, it was possible to provide additional insight into how machine learning algorithms can be fine-tuned to provide more appropriate models.

From our tests we determined that while the kNN, BDT, RF, and ERT methods all seemed to perform very similarly, obtaining a good result for the MSE $< 0.0046$, it was the RF which achieved the best metrics, and was also one of the faster algorithms in inference. However, depending on which area of the pipeline an experiment requires to be faster, the RF method could also be inefficient as it scaled worse than all other algorithms in training. Hence for problems which require regular retraining of models on large datasets one of the other algorithms such as the BDT or ERT could allow for a greater improvement. As large sky surveys producing enormous datasets will require the most efficient methods possible it could also be necessary to investigate the use of deep learning neural networks which could benefit the most when using even larger amounts of data with more features.

Further work could be done to include a wider range of machine learning algorithms, including more deep learning networks, and to test them on larger simulated datasets to confirm their scaling. By making use of the time-considered optimisation it would also be possible to further examine the trade-offs between minimising errors and the training/inference times in each individual algorithm. We could also run the benchmarks on a variety of computer architectures, making use of GPUs which have the potential to speed up the algorithms that are most parallisable, as well as allowing us to examine the environmental impact of running such computationally expensive tasks.

\section*{Acknowledgements}

B.H. was supported by the STFC UCL Centre for Doctoral Training in Data Intensive Science (grant No. ST/P006736/1). 
Authors also acknowledge the support from following grants:  O.L.'s European Research Council Advanced Grant (TESTDE FP7/291329), STFC Consolidated Grants (ST/M001334/1 and ST/R000476/1),
J.T.'s UKRI Strategic Priorities Fund (EP/T001569/1), particularly the AI for Science theme in that grant and the Alan Turing Institute, Benchmarking for AI for Science at Exascale (BASE), EPSRC ExCALIBUR Phase I Grant (EP/V001310/1).

Funding for SDSS-III has been provided by the Alfred P. Sloan Foundation, the Participating Institutions, the National Science Foundation, and the U.S. Department of Energy Office of Science. The SDSS-III web site is http://www.sdss3.org/.

SDSS-III is managed by the Astrophysical Research Consortium for the Participating Institutions of the SDSS-III Collaboration including the University of Arizona, the Brazilian Participation Group, Brookhaven National Laboratory, Carnegie Mellon University, University of Florida, the French Participation Group, the German Participation Group, Harvard University, the Instituto de Astrofisica de Canarias, the Michigan State/Notre Dame/JINA Participation Group, Johns Hopkins University, Lawrence Berkeley National Laboratory, Max Planck Institute for Astrophysics, Max Planck Institute for Extraterrestrial Physics, New Mexico State University, New York University, Ohio State University, Pennsylvania State University, University of Portsmouth, Princeton University, the Spanish Participation Group, University of Tokyo, University of Utah, Vanderbilt University, University of Virginia, University of Washington, and Yale University.

\section*{Data Availability}

The data used in this paper came entirely from the Sloan Digital Sky 
Survey data release 12 (SDSS-DR12), and is openly available from: \url{https://www.sdss.org/dr12/}.



\bibliographystyle{mnras}
\bibliography{bib}







\bsp	
\label{lastpage}
\end{document}